\pdfoutput=1
\documentclass[sigconf, natbib=true, anonymous=false]{acmart}
\usepackage{booktabs} 
\usepackage{graphicx}
\usepackage{multirow}
\usepackage[inline]{enumitem}
\usepackage{hhline}
\usepackage{setspace}
\usepackage{makecell}
\usepackage{soul}
\usepackage{xcolor,colortbl}
\usepackage{threeparttable}
\usepackage{balance}
\usepackage{float}
\usepackage{threeparttable}
\usepackage[normalem]{ulem}
\usepackage{wrapfig}
\usepackage{arydshln}
\usepackage{subfig}
\usepackage{tabularray}
\usepackage[ruled,vlined]{algorithm2e}
\newcommand{\tabincell}[2]{\begin{tabular}{@{}#1@{}}#2\end{tabular}}

\newcommand{\ie}{\emph{i.e.,}\xspace}

\AtBeginDocument{%
  \providecommand\BibTeX{{%
    \normalfont B\kern-0.5em{\scshape i\kern-0.25em b}\kern-0.8em\TeX}}}

\setcopyright{acmcopyright}
\copyrightyear{2023}
\acmYear{2023}
\acmDOI{10.1145/1122445.1122456}

\acmConference[SIGIR '24]{SIGIR '24}{July 14--18 2024}{Washington D.C.}
\acmBooktitle{SIGIR '24, July 14--18 2024, USA}
\acmPrice{15.00}
\acmISBN{978-1-4503-XXXX-X/18/06}

\renewcommand\footnotetextcopyrightpermission[1]{}

\begin{document}

\title[Aiming at the Target]{Aiming at the Target: Filter Collaborative Information \\for Cross-Domain Recommendation}

\author{Hanyu Li}
\email{hanyu-li23@mails.tsinghua.edu.cn}
\affiliation{%
  \institution{Tsinghua University}
  \city{Beijing}
  \country{China}
}
\author{Weizhi Ma}
\email{mawz12@hotmail.com}
\affiliation{%
  \institution{Tsinghua University}
  \city{Beijing}
  \country{China}
}
\author{Peijie Sun}
\email{sun.hfut@gmail.com}
\affiliation{%
  \institution{Tsinghua University}
  \city{Beijing}
  \country{China}
}
\author{Jiayu Li}
\email{jy-li20@mails.tsinghua.edu.cn}
\affiliation{%
  \institution{Tsinghua University}
  \city{Beijing}
  \country{China}
}
\author{Cunxiang Yin}
\email{jasonyin@tencent.com}
\affiliation{%
  \institution{Tencent}
  \city{Beijing}
  \country{China}
}
\author{Yancheng He}
\email{collihe@tencent.com}
\affiliation{%
  \institution{Tencent}
  \city{Beijing}
  \country{China}
}
\author{Guoqiang Xu}
\email{chybotxu@tencent.com}
\affiliation{%
  \institution{Tencent}
  \city{Beijing}
  \country{China}
}
\author{Min Zhang}
\email{z-m@tsinghua.edu.cn}
\affiliation{%
  \institution{Tsinghua University}
  \city{Beijing}
  \country{China}
}
\author{Shaoping Ma}
\email{msp@tsinghua.edu.cn}
\affiliation{%
  \institution{Tsinghua University}
  \city{Beijing}
  \country{China}
}

\begin{abstract}
As recommender systems become pervasive in various scenarios,
cross-domain recommenders (CDR) are proposed to enhance the performance of one target domain by data from other related source domains.
However, irrelevant information from the source domain may instead degrade target domain performance, which is known as the negative transfer problem. 
There have been some attempts to address this problem, mostly by designing adaptive representations for overlapped users. 
Whereas, these methods rely on the learned representations of the model, lacking explicit constraint to filter irrelevant source-domain collaborative information for the target domain, which limits their cross-domain transfer capability.

In this paper, we propose a novel \textbf{C}ollaborative information regularized \textbf{U}ser \textbf{T}ransformation (\textbf{CUT}) framework to tackle the negative transfer problem by directly filtering users' collaborative information.
In CUT, target domain user similarity is adopted as a constraint for user transformation to filter user collaborative information from the source domain.
First, CUT learns user similarity relationships from the target domain.
Then, source-target information transfer is guided by the user similarity, where we design a user transformation layer to learn target-domain user representations and a contrastive loss to supervise the user collaborative information transferring.
As a flexible and lightweight framework, CUT can be applied with various single-domain recommender systems as the backbone and extend them to multi-domain tasks.
We conduct extensive experiments of CUT with two single-domain backbone recommenders on six CDR tasks from two real-world datasets.
The results show significant performance improvement of CUT compared with SOTA single and cross-domain methods.
Further analysis illustrates that CUT can effectively alleviate the negative transfer problem.
\end{abstract}



\maketitle
\begin{figure}[ht]
    \centering
    \includegraphics[width=3.5in]{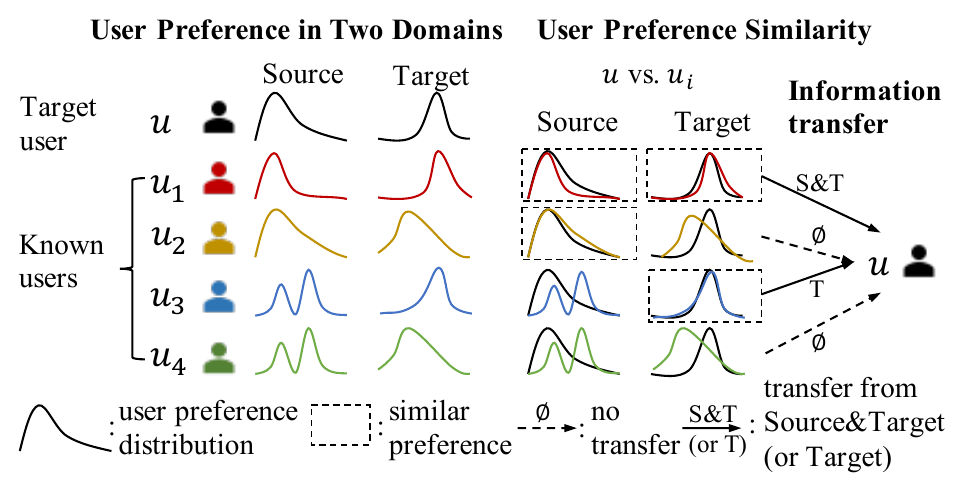}
    \caption{
  User similarity distortion in cross-domain recommendation. 
    For a target user $u$, some users have similar source-domain preferences and different target-domain preferences, e.g. $u_2$, which will lead to negative transfer of $u_2$ for $u$.
    Our CUT framework alleviates it by filtering misleading collaborative information from users with different target-domain preferences (i.e., $u_2$ and $u_4$).
    }
    \label{fig:motiv}
    \vspace{-0.5cm}
\end{figure}

\section{Introduction}
As recommender systems have been extensively applied in various scenarios, users' interaction data in multiple domains has been generated, such as different categories on e-commerce websites and different media forms on multimedia platforms.
These cross-domain interaction data enable the systems to improve recommendation accuracy in one domain~(\ie the target domain) with rich information from other domains~(\ie the source domains) including users' collaborative information, categorical preference, etc.

Various methods have been proposed for the CDR tasks, which explored
how to encode user/item representations from different domains and how to transfer information among domains. 
For instance, early attempts \cite{singh_relational_2008,hu_conet_2018,liu_cross_2020} leveraged shared embedding or network parameters to represent and transfer information in different domains. 
Another branch of cross-domain recommendation methods sought to bridge the domain gap by mapping the embeddings of users/items that are shared across domains \cite{man_cross-domain_2017,Kang2019SemiSupervisedLF,Zhu2021PersonalizedTO}. Further research modified the mapping mechanism and introduced various types of supervision signals including adversarial training and contrastive learning, to guide the mapping process~\cite{hao_adversarial_2021, li_one_2022}.
 
Although these studies are technically sound and have achieved promising results, most of them have not considered the negative transfer problem \cite{zang_survey_2022,Zhang2020ASO}. 
As information in source domain~(s) is not always useful for the target domain, adding source domain data indistinguishably during the training process may bring negative effects, which will cause the negative transfer issue for CDR.
Some recent works attempted to avoid negative transfer by designing adaptive representations for overlapped users. For example, DisenCDR~\cite{cao_disencdr_2022} learns disentangled embeddings for domain-relevant and irrelevant components, and CATART \cite{li_one_2022} proposes the attention-based representation transfer module.

However, these methods 
solely rely on the representation capacity of the model to alleviate the contradictions between domains, without attempting to fundamentally filter contradictory information.
These contradictions mainly consist of the distortion of user collaborative relationships between domains.
Particularly, users who show similar preferences in the source domain may have different interests in the target domain. 
Therefore, some collaborative knowledge in the source domain will become irrelevant or even noisy information in the target domain.
As illustrated in Figure~\ref{fig:motiv}, suppose we aim to learn the user profile of user $u$ in the target domain from known users with similar preferences. The similarity between $u$ and $u_2$ in the source domain conflicts with the target domain. This kind of irrelevant information confounds the model and induces sub-optimal performance.
Single-domain approaches utilize the target-domain information from similar users $u_1$ and $u_3$. Most previous cross-domain approaches additionally consider source-domain information from $u_1$ and $u_2$. However, the irrelevant source information of $u_2$ should be filtered out during model training.

In this paper, we propose a flexible and lightweight Collaborative information regularized User Transformation~(CUT) framework to alleviate the negative transfer problem by regularizing the similarity of user pairs, which filters irrelevant source domain collaborative information directly. In this way, introducing source domain information will not affect the integrity of the target domain user relationship. 
In CUT, we utilize single-domain recommendation methods as the backbone model and design a contrastive regularization loss term that forces the backbone models to retain the user similarity information for the target domain. By this means, CUT filters the irrelevant collaborative information from the source domain with explicit regulation signals.
Moreover, we propose a user transformation module to depict different behavior across domains of overlapping users.
Our proposed CUT framework can be seamlessly applied to various single-domain backbone models without modifying their model structure and loss terms. CUT extends them to cross-domain tasks with insignificant additional parameters and training costs.

We conduct extensive experiments in six cross-domain tasks in two real-world datasets to compare the CUT framework with state-of-the-art single-domain and cross-domain baselines. 
The results show that CUT-enhanced single-domain backbones achieve significantly better results on cross-domain tasks compared with SOTA cross-domain baseline methods.
Our main contributions can be summarized as follows:
\begin{itemize}[nolistsep, leftmargin=*]
    \item We provide a new perspective on how negative transfer occurs in cross-domain recommendations, i.e., the distortion of user similarity relationships. The \textbf{C}ollaborative information regularized \textbf{U}ser \textbf{T}ransformation~(\textbf{CUT}) framework is proposed to alleviate negative transfer by adding constraints to filter irrelevant user collaborative information. 
   \item To our knowledge, CUT is the first CDR framework that extends single-domain recommenders for multi-domain tasks without requiring modification to the backbone model structure and loss term. Thus, SOTA recommendation algorithms can be adapted for cross-domain tasks easily.
    \item Extensive experiments on cross-domain tasks show significant improvements of CUT-enhanced single-domain backbones over SOTA cross-domain and single-domain models. The source code is anonymously released in the link below \footnote{https://anonymous.4open.science/r/CUT$\_$anonymous-9815}.
\end{itemize}

\section{Related Work}
\subsection{Cross Domain Recommendation}
Cross-domain recommendation~(CDR) is proposed to enhance the performance of the target domain by utilizing data from other related domains.
Recent surveys \cite{zhu_cross-domain_2021,zang_survey_2022} divide previous works by the pattern of user/item overlap and inter/intra domain task.

When users and items have no overlap between two domains, researchers mainly rely on extracting implicit cluster-level patterns \cite{Li2009CanMA, Gao2013CrossDomainRV, Shu2018CrossFireCM} or explicit tag correlations \cite{Liu2022CollaborativeFW, Enrich2013ColdStartMW} to exploit the interaction data from the source domain and capture cross-domain analogy. More previous studies focus on partial overlap scenarios where the shared user/items can bridge the gap between domains. Pioneering studies include CMF \cite{singh_relational_2008}, where shared users naturally share the same embedding in the interaction matrix for both domains, and matrix factorization is performed with different weights for source and target domains. To model different behavior for the overlapped users across domains and transfer collaborative information between domains, models including DTCDR \cite{Zhu2019DTCDRAF} and its variations \cite{Zhu2020AGA,Zhu2021AUF} combine the representations for overlapped users in both domains for prediction. Another way to transfer knowledge between domains utilize the shared deep layers instead of separate representations. Typical methods under this paradigm include Conet \cite{hu_conet_2018}, DDTCTR \cite{Li2019DDTCDRDD}, and BiTGCF \cite{liu_cross_2020}. The widely used embedding and mapping paradigm includes EMCDR \cite{man_cross-domain_2017}, SSCDR \cite{Kang2019SemiSupervisedLF}, and PTUPCDR \cite{Zhu2021PersonalizedTO}, where the model first learns the latent factor representation for each domain, and then train a mapping function to establish the relationships between the latent space of domains.

Recent cross-domain recommenders have developed more complicated ways to fuse information from different domains. Most of the proposed models have well-motivated network structures and training techniques that organically utilize all three mentioned techniques: embedding combinations, shared deep layers, and embedding transformation for different domains. For example, AFT \cite{hao_adversarial_2021} learns the feature translations across domains under a generative adversarial network, and C$^{2}$DSR \cite{cao_contrastive_2022} utilizes the contrastive infomax objective to enhance the correlation between specific and shared user embeddings. In addition, the well-proven attention mechanism is commonly used to generate domain-specific representations from trainable global user embeddings \cite{hao_adversarial_2021,jiang_adaptive_2022,li_one_2022,cao_towards_2023}.

Most of these cross-domain recommenders rely on the overlapped users or items to directly transfer inter-domain knowledge, based on the assumption
that overlap users/items have similar collaborative relationships in the source and target domain. 
However, as illustrated in Figure~\ref{fig:motiv}, this assumption is not always true. We introduce a contrastive loss term for overlap users to transfer useful source domain knowledge while keeping the user similarity relationships in the target domain.

\subsection{Negative Transfer Problem}
Cross-domain recommenders aim to enhance the performance of the target domain using knowledge from the source domain. 
Nonetheless, not all source knowledge is useful, some of which may even contain noise and hence undermine the performance. 
Indiscriminately transferring knowledge between domains will cause the `negative transfer' problem \cite{Zhang2020ASO}.

Previous researchers have made several attempts to transfer useful knowledge while avoiding negative transfer. CDRIB \cite{cao_cross-domain_2022} encourages domain-shared information and limits domain-specific information through variational information bottleneck regularizers. DisenCDR \cite{cao_disencdr_2022} also utilizes variational auto-encoders to achieve robust performance against the irrelevant data and obtain robust domain invariant user embeddings. The domain-specific attention mechanism can also alleviate the negative transfer problem. CATART \cite{li_one_2022} relies on the attention-based representation transfer to retain useful knowledge from other domains under the data isolation constraint. The recent UniCDR \cite{cao_towards_2023} which tackles multiple CDR tasks simultaneously addresses this problem with interaction level and domain level masking and contrastive loss to obtain robust domain-shared user representation. We notice that most previous models address the negative transfer problem using different forms of single-user representations for different domains, but they fail to model the inter-user similarity shift across domains.

These cross-domain recommenders can partially mitigate the negative transfer problem.
However, all these approaches only focus on adapting the representations of single users to fit in the target domain, neglecting the differences in collaborative information between users in source and target domains.
Our proposed CUT framework explicitly introduces constraints to retain the user similarity relationships in the target domain, which directly filters the negative transfer information. Furthermore, most existing cross-domain recommenders have fixed model structures,
which require additional effort to train and apply.
In contrast, our framework can extend various single-domain recommenders to cross-domain tasks while keeping their model structure and loss terms. 

\begin{table}[t]
  \caption{Primary notations used in this paper.}
  \label{tab:not}
  \small
  \begin{tabular}{ll}
    \toprule
    \textbf{Notations} & \textbf{Definitions} \\
    \midrule
    $\mathcal{S}=(\mathcal{U}^s,\mathcal{I}^s,\mathbf{M}^s)$ & Source domain dataset.\\
    $\mathcal{U}^s = \{u_1^s,...,u_k^s\}$ & Source user set. \\
    $\mathcal{I}^s= \{i_1^s,...,u_j^s\}$ & Source item set.\\
    $\mathbf{M}^s \in \{0,1\}^{k\times j}$ & Source binary interaction matrix.\\
    \midrule
    $\mathcal{T}=(\mathcal{U}^t,\mathcal{I}^t,\mathbf{M}^t)$ & Target domain dataset.\\
    $\mathcal{U}^t = \{u_1^t,...,u_n^t\}$ & Target user set. \\
    $\mathcal{I}^t= \{i_1^s,...,u_m^t\}$ & Target item set.\\
    $\mathbf{M}^t \in \{0,1\}^{n\times m}$ & Target binary interaction matrix.\\
    \midrule
    \tabincell{l}{$\mathcal{U}^o = \mathcal{U}^s \cap \mathcal{U}^t $} & Overlapped Users.\\
    \midrule
    \midrule
    $R_1(u,i)$, $\Phi_1$ & \tabincell{l}{Single-domain backbone model for\\phase TARGET with parameters $\Phi_1$.}\\
    $\Theta_{t_1}$ & \tabincell{l}{The embedding parameters of\\target users in phase TARGET.}\\
    \midrule
    $R_2(u,i)$, $\Phi_2$ & \tabincell{l}{Single-domain backbone model for\\phase TRANSFER with parameters $\Phi_2$.}\\
    $\Theta=\{\Theta_{t}$, $\Theta_{o}$, $\Theta_{s}\}$ & \tabincell{l}{The embedding parameters of target-\\only,  overlapped, and source-only\\users in phase TRANSFER.}\\
    $\mathbf{u}_p^{t\Theta_{t}}$ & \tabincell{l}{User embedding of target user $p$\\in phase TRANSFER.}\\
    \midrule
    $\mathbf{L_s},\mathbf{L_t},\mathbf{L_c}$ & \tabincell{l}{Loss term of source, target and\\contrastive  regularization , respectively.}\\
    $\mathbf{A^\gamma} \in \{0,1\}^{n\times n}$ & User similarity matrix with threshold $\gamma$.\\
    $\mathbf{F}$ & User transformation layer.\\
    \bottomrule
  \end{tabular}
\end{table}

\begin{figure*}[t]
    \centering
    \includegraphics[width=\linewidth]{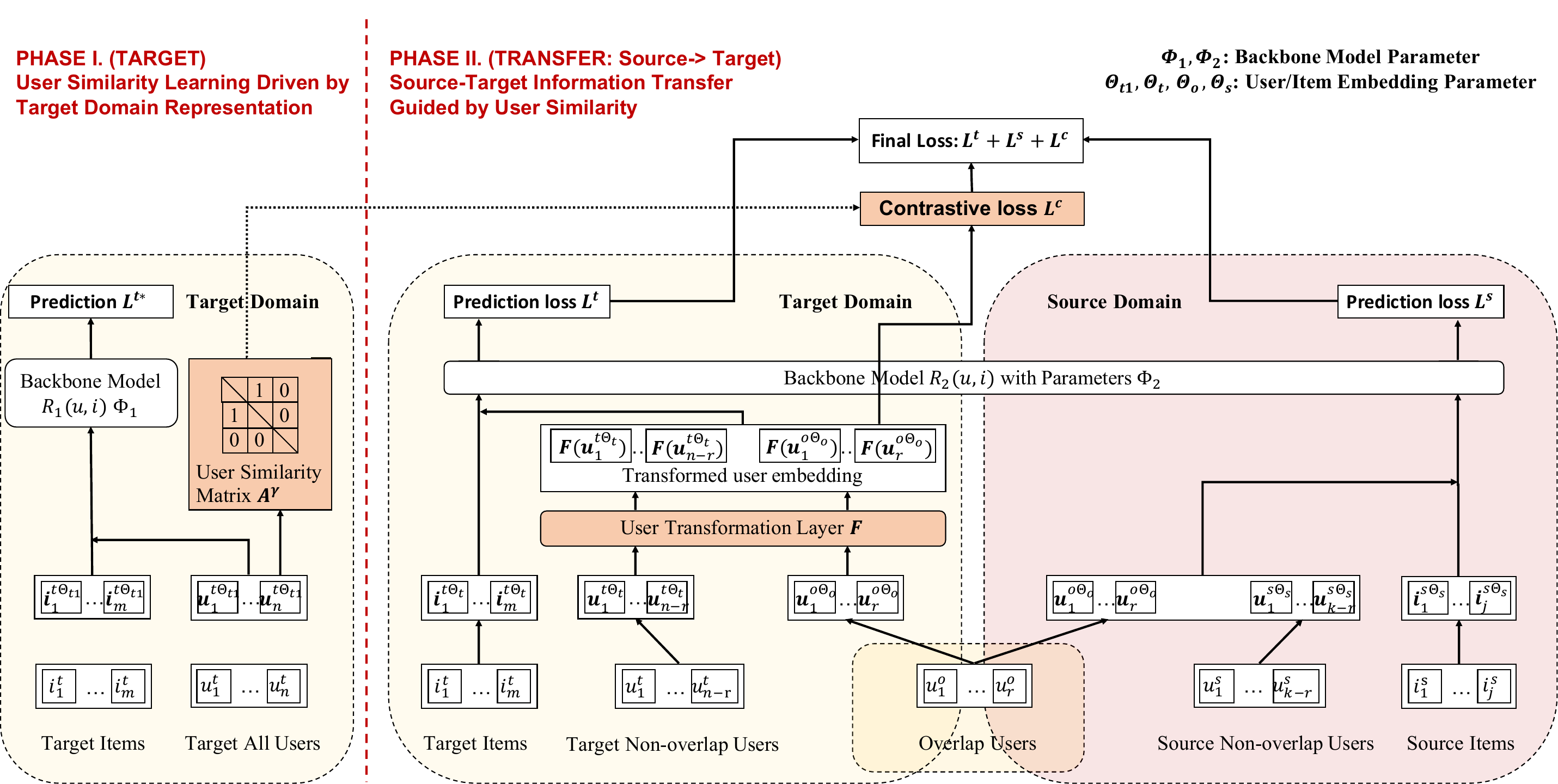}
    \caption{An overview of Collaborative information regularized User Transformation~(CUT) framework. It includes a TARGET phase to learn user similarity and a TRANSFER phase to filter useful source information to transfer to the target domain. Key components of the TRANSFER phase include a user transformation layer and a specially designed contrastive loss.
    }
    \label{fig:basemodel}
\end{figure*}
\section{CUT Framework}
\subsection{Framework Overview}
Our notation in this paper is defined in Table \ref{tab:not}. 
In the intra-domain CDR task, we intend to improve the recommendation performance on target domain $\mathcal{T}$ 
with training data from both source and target domains, $\mathbf{M}^s$ and $\mathbf{M}^t$. 

Our proposed Collaborative information regularized User Transformation~(CUT) framework is shown in Figure~\ref{fig:basemodel}.
In CUT, any single-domain backbone recommender system $R(u,i)$ with user and item embedding layers can be applied for the cross-domain task. 
We denote its parameter set as $\Phi$. The embedding layer parameters are expressed as $\Theta$. 

The training procedure of CUT includes two phases: First, a user similarity learning~(TARGET) phase driven by target domain representation is adopted to obtain the user collaborative information in the target domain, i.e., a binary target user similarity matrix $\mathbf{A} \in \{0,1\}^{n\times n}$.
At the TARGET phase, a backbone model $R_1(u,i)$ with parameters $\Phi_1$ and embedding parameters $\Theta_{t1}$ is trained in the target domain. Details of this phase will be described in Section~\ref{seq:sim}.
Second, a source-target information transfer (TRANSFER) phase guided by user similarity is designed to transfer the information from the source domain while avoiding the negative transfer problem.
TRANSFER is the core phase in CUT, which includes a backbone model $R_2(u,i)$ with the same structure as the TARGET phase but different parameters $\Phi_2$. 
$R_2(u,i)$ is trained on all data from both the source and target domains, and
the target-only, overlap, and source-only embedding parameters are $\Theta_{t}$, $\Theta_{o}$, and $\Theta_{s}$, respectively.
We design two components to ensure that $R_2(u,i)$ learns proper representations of target domain users and useful cross-domain information from the source domain:
a user representation transformation layer to 
model user representations in the target domain~(Section~\ref{sec:rep_trans}), and a contrastive negative transfer regularization loss term $\mathbf{L}^c$ to retain the user similarity relationships of matrix $\mathbf{A}$~(Section~\ref{sec:contra_loss}).
We will explain why this procedure alleviates negative transfer.
Finally, we illustrate the model-agnostic training pipeline, which can extend arbitrary single-domain recommendation backbone $R(u,i)$ to cross-domain tasks~(Section~\ref{sec:cut_training}).

\subsection{User Similarity Learning Driven by Target Domain Representation}
\label{seq:sim}
In the TARGET phase, we aim to obtain the user collaborative information~(i.e., pair-wise similarity) in the target domain. The similarity will work as an explicit supervision signal upon the final user embedding in the TRANSFER phase. 
We define a binary user similarity matrix $\mathbf{A}\in \{0,1\}^{n\times n}$, where $A_{i,j}=1$ denotes user $u_i$ and $u_j$ are similar.
A binary, rather than continuous, measure of similarity is adopted because the similarity is only an estimation of the collaborative relationship between users, and overly strict estimation would induce unnecessary or even erroneous constraints.
Let $\phi(\mathbf{a},\mathbf{b})=\frac{\mathbf{a}\cdot\mathbf{b}^{T}}{|\mathbf{a}||\mathbf{b}|}$ be the cosine similarity between two vectors $\mathbf{a}$ and $\mathbf{b}$,
and $\gamma$ be the similarity threshold as a hyper-parameter, we have two approaches to acquire the binary similarity between users,
\begin{align}
    \mathbf{A'}^{\gamma}_{pq} &= \mathbb{I}(\phi(\mathbf{M}^t_{p},\mathbf{M}^t_{q})>\gamma),\qquad p,q \in \{1,...,n\}\label{eq:his_sim}\\
    \mathbf{A}^{\gamma}_{pq} &= \mathbb{I}(\phi(\mathbf{u}_p^{t\Theta_{t_1}},\mathbf{u}_q^{t\Theta_{t_1}})>\gamma),\qquad p,q \in \{1,...,n\}\label{eq:train_sim}
\end{align}
The similarity in Equation \ref{eq:his_sim} is based on the interaction history of the users in the target domain. Equation \ref{eq:train_sim} is based on the user representations of the single-domain backbone $R_1(u,i)$ with parameters $\Theta_{t_1}$ which is trained on the target set in the TARGET phase as shown in Figure \ref{fig:basemodel}. 
We argue that the latter approach yields finer calibrated similarities between users because $\mathbf{A}^{\gamma}$ represents the exact collaborative relationship of users that backbone model $R_1(u,i)$ extracts from the target domain. Whereas, $\mathbf{A'}^{\gamma}$ fails to consider the backbone model $\Theta_{t1}$. For instance, it yields the same similarity relationships for a simple backbone MF and for a stronger backbone LightGCN, which will lead to an imprecise supervision signal for the following domain-transfer phase. 
Therefore, we choose the similarity generated by $R_1(u,i)$ as in Equation~\ref{eq:train_sim}. 
Empirical results in Section~\ref{sec:exp_similarity} will confirm the rationality of this choice.

\subsection{User Representation Transformation Layer}
\label{sec:rep_trans}
In the TRANSFER phase, source and target domains share the same backbone model parameter $\Phi_2$ to exchange cross-domain information, where the overlapping users should have different representations in source and target domains to indicate the domain differences. 
However, 
simply assigning two distinct sets of embeddings for overlapping users overlooks the sharing of information for the same users between domains.

To solve this issue, we introduce a User Representation Transformation module to depict the relationship between the source and target domains of the overlapping users.
To be specific, for an overlapping user $u_i^o$, we directly learn its source-domain embedding $\mathbf{u}_i^{o\Theta_o}$ by the backbone model and adopt a transformation layer $\mathbf{F}$ to learn its target-domain embedding $\mathbf{F}(\mathbf{u}_i^{o\Theta_o})$.
Note that 
we also transfer the non-overlap target users $u^t$ to ensure their embeddings under the same distribution.
For simplicity, we follow the efficient one-layer MLP structure for user representation transformation.
The user transformation layer is activated when input user $u^t$, $u^o$ with embedding $\mathbf{u}^{t\Theta_t}, \mathbf{u}^{o\Theta_o}$ are from the target domain:
\begin{align}
    \mathbf{F}(\mathbf{u}^{t\Theta_t}) &= MLP(\mathbf{u}^{t\Theta_t}) = \mathbf{W} \cdot \mathbf{u}^{t\Theta_t} +\mathbf{b}\\
    \mathbf{F}(\mathbf{u}^{o\Theta_o}) &= MLP(\mathbf{u}^{o\Theta_o}) = \mathbf{W} \cdot \mathbf{u}^{o\Theta_o} +\mathbf{b}
\end{align}
Where $\mathbf{W}$ and $\mathbf{b}$ are learnable parameters.
Afterwards, the transferred target-domain user embeddings will be fed into backbone $R_2(u,i)$, and share parameters with the source-domain interactions.
The adaptation module works as an identifier of the domain, which allows CUT to model different user behaviors across domains while keeping useful domain-shared knowledge of users.
Note that no transformation is adopted for the item embeddings since items in source and target domains lie in different spaces naturally.

\subsection{Contrastive Negative Transfer Regularization Loss}
\label{sec:contra_loss}
According to our experiments, the negative transfer severely degrades the performance when single-domain approaches are directly applied to cross-domain scenarios. 
This issue is partially addressed by the above user transformation layer since it considers the differences of overlapping users across domains.
However, only modeling single-user transformation is not enough, since it is still under the basic assumption that overlapped users share similarity relationships across domains, which may still cause negative transfer as we illustrated in Figure~\ref{fig:motiv}.

To further eliminate the negative transfer issue, we propose an additional loss term based on the supervised contrastive learning mechanism \cite{Khosla2020SupervisedCL} to regularize the collaborative information~(i.e., similarity) of user representations.
Because we only focus on the performance of the target domain, the collaborative information in the target domain is essential.
Thus, the goal of our contrastive regularization loss term is to ensure that the user similarity relationships of the target domain users will not drift away under the incoming source domain information.
This loss term mainly guides the training process of the user transformation layer $\mathbf{F}$, which is also only activated for the target-domain users $\mathcal{U}^t$.
Specifically, for each target mini-batch $b\in B^t$ with user set $U_b$, we extract target user pairs (including target and overlapping users) according to the binary target user similarity matrix $\mathbf{A}^{\gamma}$, 

\begin{align}
    S_b &= \{(u^t_i,u^t_j)|(u^t_i,u^t_j\in U_b,i\neq j) \wedge (\mathbf{A}^{\gamma}_{ij}=1)\}\\
    A_b &= \{(u^t_x,u^t_y)|u^t_x,u^t_y\in U_b,x\neq y\}
\end{align}
Where $S_b$ denotes similar user pairs, and $A_b$ denotes all user pairs.
Afterward, we optimize a contrastive loss $\mathbf{L}^c$ to guide the transformed target user representations to retain the user similarity relationships of the original target domain information,
\begin{equation}
\small
\label{eq:L_c}
    \mathbf{L}^c=\sum_{b \in B^t}-\frac{1}{|S_b|} \sum_{(u^t_i,u^t_j)\in S_b}\log \frac{|A_b|\exp \left(\mathbf{F}(\mathbf{u}^{t\Theta_t}_i) \cdot \mathbf{F}(\mathbf{u}^{t\Theta_t}_j) / \tau\right)}{\sum_{(u^t_x,u^t_y)\in A_b}\exp \left(\mathbf{F}(\mathbf{u}^{t\Theta_t}_x) \cdot \mathbf{F}(\mathbf{u}^{t\Theta_t}_y) / \tau\right)}
\end{equation}
Where temperature $\tau$ is a hyper-parameter.
$\mathbf{L}^c$ in Equation \ref{eq:L_c} decreases when originally similar users in the target domain are still similar after being transformed to the source domain space, while increases when the similarity relationship is changed after transformation. 
This contrastive negative transfer regularization loss term guarantees that although the source domain version representation of overlapped user $u^o$, i.e. $\mathbf{u}^{o\Theta_o}$, involves in the source domain training process, the transformed target domain version user representation $\mathbf{F}(\mathbf{u}^{o\Theta_o})$ and $\mathbf{F}(\mathbf{u}^{t\Theta_t})$ will retain its similarity relationships in the target domain.
Note that the mini-batch may contain repeating user IDs (i.e. when $u^t_i=u^t_j$). We filtered these user pairs out of $S_b$ and all user pair sets, because these same-user pairs do not provide meaningful supervision signals.

\subsection{Training Process for CUT}
\label{sec:cut_training}
\begin{algorithm}[t]
	\caption{Training Process of Collaborative information regularized User Transformation~(CUT).}
	\label{alg:algorithm1}
        \textbf{TARGET PHASE}
        
	\KwIn{Target user set $\mathcal{U}^t$, Target item set $\mathcal{I}^t$, Target interaction matrix $\mathbf{M}^t$, Similarity threshold $\gamma$}
	\KwOut{Target binary user similarity matrix $\mathbf{A}^{\gamma}$}  
	Train a backbone model $R_1(u,i)$ with $\Phi_1$ on the target domain $(\mathcal{U}^t,\mathcal{I}^t,\mathbf{M}^t)$ with loss $\mathbf{L}^{t*}$;\\
        Derive target user similarity matrix $\mathbf{A}^{\gamma}$ by Equation \ref{eq:train_sim};

        \BlankLine
        
        \textbf{TRANSFER PHASE}
        
	\KwIn{Source \& target domain data $(\mathcal{U}^s,\mathcal{I}^s,\mathbf{M}^s)$, $(\mathcal{U}^t,\mathcal{I}^t,\mathbf{M}^t)$, Hyper-parameters $\alpha, \lambda$}
	\KwOut{Backbone model parameters $\Phi_2$; User transformation Layer $\mathbf{F}$; User \& item embeddings $\mathbf{u}^{\Theta}$, $\mathbf{i}^{\Theta}$}  
	Train a backbone model $R_2(u,i)$ on both domain with contrastive loss term, target prediction loss and source prediction loss $\mathbf{L}^{all}=(1-\alpha)\mathbf{L}^{t}+\alpha\mathbf{L}^{s}+\lambda\mathbf{L}^{c}$;
\end{algorithm}

As illustrated in Algorithm \ref{alg:algorithm1}, the training process is divided into two phases: TARGET and TRANSFER. 
The TARGET phase outputs the target binary user similarity matrix $\mathbf{A}^{\gamma}$.
For any single-domain recommendation model $R(u,i)$, we first train a backbone model $R_1(u,i)$ with parameter $\Phi_1$ on the target domain to get the target user embeddings $\mathbf{u}^{t\Theta_{t1}}$ with parameters $\Theta_{t1}$.
Then the binary similarity matrix $\mathbf{A}^{\gamma}$ is derived from Equation~\ref{eq:train_sim}. 
The TRANSFER phase induces useful source-domain knowledge into the model with the user transformation layer and the contrastive negative transfer regularization loss.
In this phase, we train another backbone model $R_2(u,i)$~(with the same structure as $R_1(u,i)$) with $\Phi_2$ on data from both source and target domains. 
During training $R_2(u,i)$, each mini-batch contains a batch of interactions from the source domain and a batch from the target domain. 
For the target domain interactions, every user embedding $\mathbf{u}^{t\Theta_t}$ will first be transformed by the user transformation layer $\mathbf{F}$, and then fed into the backbone model. 
These two batches yield the prediction losses on the source domain $\mathbf{L}^s$ and target domain $\mathbf{L}^t$ respectively. 
The form of prediction loss depends on the backbone model. 
The final loss term consists of three terms as in Equation \ref{eq:loss}: source domain prediction loss $\mathbf{L}^s$, target domain prediction loss $\mathbf{L}^t$, and contrastive loss $\mathbf{L}^c$ from Equation~\ref{eq:L_c}.

\begin{equation}
\label{eq:loss}
    \mathbf{L}^{\text{all}} = (1-\alpha)\mathbf{L}^{t}+\alpha\mathbf{L}^{s}+\lambda\mathbf{L}^{c}
\end{equation}
Where $\alpha$ and $\lambda$ are both hyper-parameters that denote the weights of each part of losses.
\section{Experiments}
In this section, we conduct extensive experiments on six cross-domain tasks from two real-world datasets and compare the performance of our CUT framework with state-of-the-art single and cross-domain recommenders on the target domain. We aim to answer the following research questions: 

\textbf{RQ1}: How well does CUT perform compared to the SOTA single and cross-domain baselines? 

\textbf{RQ2}: Does CUT alleviate the negative transfer problem? 

\textbf{RQ3}: How does target-driven user similarity take effect on cross-domain recommendation performance? 

\subsection{Experimental Settings}
\subsubsection{Datasets \& Evaluation Metrics}
We use three pairs of domains on two real-world cross-domain datasets to evaluate the performance of the CUT framework and other baselines. Each one in the domain pairs is treated as the target domain for a cross-domain task, adding up to six cross-domain tasks in all.
\begin{itemize}
    \item {\textbf{Amazon dataset}\footnote{http://jmcauley.ucsd.edu/data/amazon/index\_2014.html}:} It is a large-scale e-commerce dataset with item interactions
    from multiple domains. We choose two pairs of domains, Cloth\&Sports, and Cloth\&Video, and perform cross-domain recommendations respectively. Cloth and Sports are more closely related while Cloth and Video share less cross-domain knowledge.
    \item {\textbf{Douban dataset}\footnote{https://recbole.s3-accelerate.amazonaws.com/CrossDomain/Douban.zip}:} Douban is a music and movie online 
    platform, where we consider two cross-domain tasks with music and movie as target/source domains, respectively.
\end{itemize}
Both datasets are widely used for cross-domain recommendation models \cite{liu_cross_2020,cao_disencdr_2022,li_one_2022,cao_towards_2023}. Following the previous work, we transform the ratings into implicit data where each entry is marked as 0 or 1 according to whether the user has interacted with the item.
We filter the dataset to keep users and items with at least 5 interactions and split the user history with the ratio of 8:1:1 for training, validation, and testing in the target domain for each user. The source domain is split by 8:2 for training and validation for fair comparisons with some previous CDR methods with source phases.
The dataset statistics are listed in Table~\ref{tab: data}.
For evaluation, a full ranking setting is utilized, where the recommendation is conducted on all items in the datasets.
We evaluate all CDR tasks by HR@10 and NDCG@10 on the target domain; both are commonly used evaluation metrics.

\begin{table}[hbt]
  \caption{Dataset statistics. The subscript $o$ indicates \textit{overlap}.}
  \label{tab: data}
  \small
  \begin{tabular}{c|c|ccccc}
\hline
\multicolumn{1}{c|}{\textbf{Dataset}}                & \textbf{Domain} & $|\mathcal{U}|$ & $|\mathcal{I}|$ & \# \textbf{Clicks} & $|\mathcal{U}_o|$ & $|\mathcal{I}_o|$ \\ 
\hline
\multirow{4}{*}{Amazon} & Sports  & 35,599   & 18,358   & 296,337   & \multirow{2}{*}{3,908}                & \multirow{2}{*}{704}                 \\ 
                        & Cloth   & 39,388   & 23,034   & 278,677   &                                      &                                      \\ 
                        \cline{2-7} 
                        & Video   & 24,034   & 10,673   & 231,780   & \multirow{2}{*}{999}                 & \multirow{2}{*}{0}                   \\ 
                        & Cloth   & 39,388   & 23,034   & 278,677   &                                      &                                      \\ 
                        \hline 
\multirow{2}{*}{Douban} & Music   & 16,041   & 40,405   & 1,140,090 & \multirow{2}{*}{14,000}               & \multirow{2}{*}{0}                   \\ 
                        & Movie   & 22,254   & 27,432   & 2,760,500 &                                      &                                      \\ 
                        \hline
\end{tabular}
\end{table}

\begin{table*}[hbt]
  \caption{Performance comparisons on six cross-domain tasks. * shows statistical significance (paired t-test with p-value < 0.05). The best performance is in bold, and the second-best results are underlined.}
  \label{tab:Experiment}
  \small
  \begin{tabular}{c|c|c|cc|cc|cccc|cc}
    \toprule
    \multirow{3}{*}[-4ex]{\textbf{Dataset}} & \multirow{3}{*}[-4ex]{\makecell{\textbf{Domain:}\\\textbf{Source}\\$\rightarrow$ \textbf{Target}}} & \multirow{3}{*}[-4ex]{\makecell{\textbf{Metrics}\\\textbf{(@10)}}}& \multicolumn{4}{c|}{\textbf{Single Domain Methods}} & \multicolumn{4}{c|}{\multirow{2}{*}[-2ex]{\textbf{Cross Domain Methods}}}& \multicolumn{2}{c}{\multirow{2}{*}[-2ex]{\textbf{Our Methods}}}\\
    \cline{4-7}
    & & &\multicolumn{2}{c|}{\makecell{Trained on\\Target Domain}}&\multicolumn{2}{c|}{\makecell{Trained on\\Both Domains}}&&&&&&\\
     \cline{4-13}
    & && MF	& LightGCN & CMF & LightGCN &EMCDR&DTCTR& CAT-ART & UniCDR & \makecell{CUT-\\MF} & \makecell{CUT-\\LightGCN}  \\
    \hline
    \multirow{8}{*}{\textbf{Amazon}}& \multirow{2}*{\makecell{\textbf{Cloth}\\ $\rightarrow$ \textbf{Sports}}} & Recall & 0.0492 & 0.0604 & 0.0545 & 0.0614 & 0.0538 & 0.0558 & 0.0515 & \uline{0.0624} & 0.0601 & \textbf{0.0653*}  \\
    
    & & NDCG & 0.0270 & 0.0331 & 0.0293 & 0.0335 & 0.0288 & 0.0332 & 0.0276 & \uline{0.0340} & 0.0335 & \textbf{0.0364*}\\
     \cline{2-13}
    & \multirow{2}*{\makecell{\textbf{Sports}\\$\rightarrow$ \textbf{Cloth}}} & Recall & 0.0243 & 0.0385 & 0.0291 & 0.0421 & 0.0234 & 0.0263 & 0.024 & \uline{0.0433} & 0.0393 & \textbf{0.0441*}\\
    & & NDCG & 0.0137 & 0.0207 & 0.0157 & 0.0231 & 0.0127 & 0.0141 & 0.0130 & \uline{0.0239} & 0.0222 & \textbf{0.0252*}\\
    \cline{2-13}
    &  \multirow{2}*{\makecell{\textbf{Cloth}\\$\rightarrow$ \textbf{Video}}} & Recall & 0.1153 & 0.1181 & 0.1194 & 0.1171 & 0.1165 & 0.1085 & 0.1133 & 0.1249 & \uline{0.1275} & \textbf{0.1303*}  \\
    
    & & NDCG & 0.0623 & 0.0639 & 0.0644 & 0.0639 & 0.0633 & 0.0584 & 0.0609 & 0.0684 & \uline{0.0704} & \textbf{0.0720*} \\
     \cline{2-13}
    & \multirow{2}*{\makecell{\textbf{Video}\\$\rightarrow$ \textbf{Cloth}}} & Recall & 0.0243 & \textbf{0.0385} & 0.0246 & 0.0379 & 0.0232 & 0.0241 & 0.0245 & 0.0349 & 0.0362 & \uline{0.0381} \\
    & & NDCG & 0.0137 & \uline{0.0207} & 0.0136 & 0.0206 & 0.0124 & 0.0126 & 0.0123 & 0.0191 & 0.0203 & \textbf{0.0213}\\
    \hline
     \hline
    \multirow{4}{*}{\textbf{Douban}} & \multirow{2}*{\makecell{\textbf{Movie}\\$\rightarrow$ \textbf{Music}}} & Recall & 0.1004 & 0.1069 & 0.0944 & 0.0972 & 0.1014 & 0.0881 & 0.0901 & 0.1073 & \textbf{0.1238*} & \uline{0.1205} \\
    & & NDCG & 0.0733 & 0.0806 & 0.0725 & 0.0772 & 0.0756 & 0.0658 & 0.0685 & 0.0754 & \textbf{0.0952*} & \uline{0.0946} \\
     \cline{2-13}
    & \multirow{2}*{\makecell{\textbf{Music}\\$\rightarrow$ \textbf{Movie}}} & Recall & 0.1053 & 0.1004 & 0.0946 & 0.0966 & 0.1064 & 0.0943 & 0.1055 & 0.1095 & \uline{0.1390} & \textbf{0.1393*} \\
    & & NDCG & 0.0997 & 0.0997 & 0.1031 & 0.1096 & 0.1156 & 0.0982 & 0.1048 & 0.0994 & \uline{0.1413} & \textbf{0.1437*}\\
    \hline
  \end{tabular}
\end{table*}

\subsubsection{Compared Baselines}
Our CUT framework is applied to two single-domain backbones and compared against the SOTA cross-domain baselines. All compared methods are listed below.

\textbf{Single-domain baselines} are trained solely on the target dataset:

\begin{itemize}
    \item{\textbf{MF} \cite{Koren2009MatrixFT}} is the classic matrix factorization model that first represents users and items with latent factors.
    \item{\textbf{LightGCN} \cite{He2020LightGCNSA}} is a well-known effective baseline for top-K recommendation that models collaborative information using a simplified graph convolutional network.
\end{itemize}

We also evaluate the performance of \textbf{directly training our single-domain backbones on both domains}, where overlapped users/items share the same embedding. 
These straightforward attempts utilize collaborative information from the source domain without filtering, neglecting the negative transfer problem.
\begin{itemize}
    \item{\textbf{CMF} \cite{singh_relational_2008}} is a often-compared classical cross-domain recommender. It is MF trained on both domains with different weights in prediction loss.
    \item{\textbf{LightGCN}} is also adopted for training on both domains. The prediction loss is also differently weighted between domains.
\end{itemize}

For \textbf{cross-domain baselines}, we compare our CUT framework with classical (EMCDR, DTCDR) and SOTA algorithms (UniCDR, CAT-ART) that consider the negative transfer problem.
\begin{itemize}
    \item{\textbf{EMCDR} \cite{man_cross-domain_2017}} first proposes the embedding and mapping framework where a mapping function is trained to project source user embeddings to the target domain.
    \item{\textbf{DTCDR} \cite{Zhu2019DTCDRAF}} combines the representation of overlapping users to learn the domain-shared knowledge.
    \item{\textbf{CAT-ART} \cite{li_one_2022}} tackles the negative transfer problem with a robust global user representation and an attention-based representation transfer module.
    \item{\textbf{UniCDR} \cite{cao_towards_2023}} transfers the most relevant domain-shared information across domains by its domain-shared and specific user embeddings and encourages the information transfer using interaction-level contrastive learning.
\end{itemize}

For \textbf{the proposed CUT framework}, we use two single-domain approaches as backbones for CUT, namely \textbf{CUT-MF} and \textbf{CUT-LightGCN}. Surprisingly, even classical single-domain backbones obtain competitive or even better performance against the latest cross-domain ones after being enhanced by our CUT framework.

\subsubsection{Implementation Details}
For a fair comparison, we conduct a grid search for the hyper-parameters for all baselines in the open-source RecBoleCDR \cite{zhao2021recbole} library.
The embedding size is fixed at 64 for all models. 
In our proposed CUT framework, the batch size is 2048, the learning rate of Adam \cite{Kingma2014AdamAM} optimizer is set as 0.001, the domain weight factor $\alpha$ is 0.2, and the cosine similarity threshold $\gamma$ is 0.9. The weight of contrastive loss term $\lambda$ is 1e-4 for the Amazon dataset and 5e-5 for the Douban dataset, which is chosen by log-scaled grid search. The weight decay is set to 1e-6 for the Amazon dataset and 1e-7 for Douban after grid search, and after comparison, we adopt binary cross entropy loss for the Amazon dataset and Bayesian Personalized Ranking (BPR) loss \cite{Rendle2009BPRBP} for Douban. As illustrated in Section \ref{sec:contra_loss}, user similarity is only needed for target users that co-exist in each mini-batch in phase TRANSFER. Therefore, in practice, our framework does not maintain the whole similarity matrix with $|U|^2$ elements. Only the fixed target user embedding from Phase TARGET is saved for calculating cosine similarity. We also track the time cost of CUT-LightGCN and LightGCN trained on both datasets for the Amazon Cloth \& Sports dataset. The mean time cost of a training epoch is 5.13s and 4.87s respectively. Our framework induces controllable time and space consumption to the single-domain backbone.
We have released the source code anonymously in the link below \footnote{https://anonymous.4open.science/r/CUT$\_$anonymous-9815}.

\subsection{Overall Performance (RQ1)}
To answer \textbf{RQ1}, we compare our CUT framework with other single-domain and state-of-the-art cross-domain baselines on six domain transfer tasks of two datasets. Performances of target domain recommendation on six tasks are shown in Table~\ref{tab:Experiment}, respectively.

According to the results, our CUT framework consistently yields better performance due to the explicit regularization of the user similarity relationships across domains.
In Amazon Cloth \& Video and Douban datasets, the relationship between domains is relatively distinct.
Therefore, our CUT framework significantly outperforms the best baseline by a large margin (15\% Recall and 18\% NDCG for Douban Movie $\rightarrow$ Music), for the CUT framework is more robust against irrelevant information from the source dataset, and it utilizes the dense source dataset better by retaining more useful knowledge because of the user transformation layer.
When the domains are complementary and relatively sparse, i.e. Amazon Sports and Cloth dataset, CUT still achieves 7\% better NDCG and 4.6\% better Recall value than the best baseline. 
Generally, CUT benefits from a stronger backbone, because a stronger backbone offers finer-grained collaborative knowledge modeling. In addition, a stronger backbone provides more accurate user similarity relationships while pre-training. LightGCN outperforms MF greatly when the dataset is relatively sparse, while on dense datasets, the performance of LightGCN and MF shares little difference. The Douban dataset is denser than the Amazon dataset. In addition, MF has a very simple model structure (the multiplication of user \& item embeddings), which means that the direct constraint on the target user embeddings in CUT contributes more to the final results. Thus, CUT-MF achieves better results in the Douban Movie \& Music dataset.

Therefore, we have \textbf{Answer to RQ1}:
The proposed CUT framework outperforms SOTA baselines significantly on most evaluation metrics on six CDR tasks from two real-world datasets.

Since CUT shows similar performance improvement on two backbones, 
we adopt LightGCN as the backbone for the CUT framework in the following analyses.

\vspace{-0cm}
\subsection{Model Performance on Sparse Target Domain Data~(RQ1)}
\begin{figure}[t]
    \centering
    \setlength{\belowcaptionskip}{-0.5cm} 
    \includegraphics[width=\linewidth]{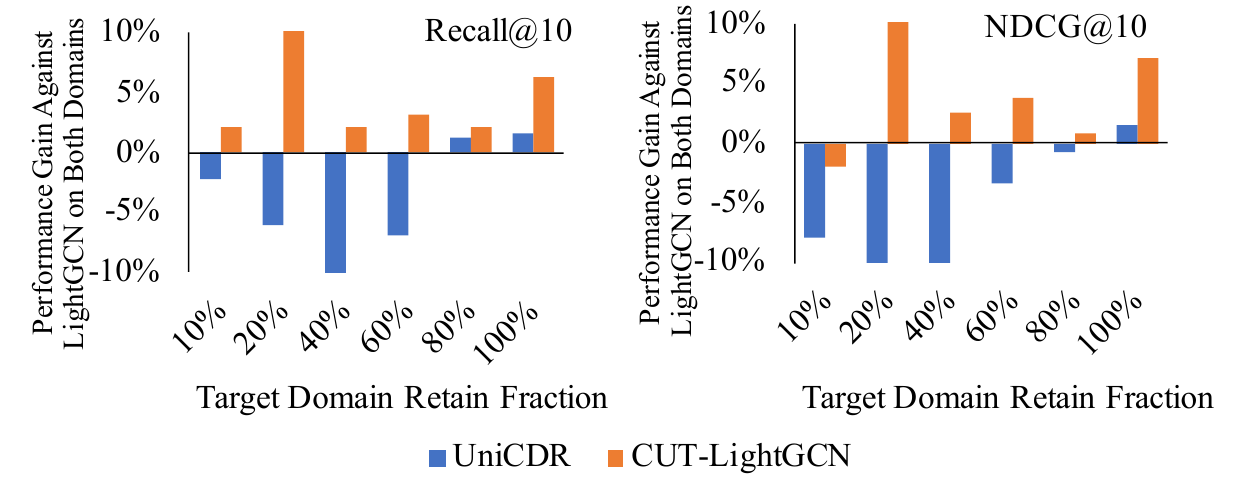}
    \caption{Performance on sparse target domain dataset. We randomly sample the target Amazon Sports training dataset interactions with different retain fractions, while fixing the source Amazon Cloth dataset and the target test dataset.}
    \label{fig:sparse}
\end{figure}

In common practice, cross-domain models often aim at transferring knowledge from domains with abundant data to sparse domains. To examine the performance of CUT and other baselines under this scenario (\textbf{RQ1}), we study the effect of different training data sizes of the target domain. While keeping other configurations fixed, we sample the target domain training set to retain different fractions of interactions. As shown in Figure \ref{fig:sparse}, our CUT framework outperforms the best baseline UniCDR and LightGCN trained on both training datasets, when we retain at least 20\% of the interactions. Note that the comparison is based on the complete target domain test set, and our user similarity regularization term is derived from the sampled target domain training set. Therefore, random sampling will undermine the performance of our CUT framework, which leads to smaller improvements in the sparse target datasets. However, even in the worst case, our regularization term does not degrade target domain performance. The reason is that the sparse target training set belongs to the same distribution as the test set, so the derived target-user similarity relationship is less informative but still unbiased.

\begin{table*}[t]
\centering
\caption{Case study on a pair of users $(u_1,u_2)$. \textbf{TagSim} represents the overlap ratio of the 5 most popular tags between the two users. \textcolor{red}{Tags with red color} in two result lists are the ones matching the tags of the target domain ground truth.
$u_1$ and $u_2$ have similar preferences on the source domain and different preferences on the target domain. 
The recommendation results from our CUT framework provide different items for $u_1$ and $u_2$ in the target domain, 
while the SOTA cross-domain model UniCDR generates similar target-domain recommendations.
}
\begin{tblr}{
  cells = {c},
  cell{1}{2} = {c=5}{},
  cell{2}{1} = {r=2}{},
  cell{4}{1} = {r=2}{},
  cell{6}{1} = {r=2}{},
  cell{8}{1} = {r=2}{},
  cell{10}{1} = {r=2}{},
  vline{2,7} = {-}{},
  hline{1,12} = {-}{0.08em},
  hline{2} = {-}{0.05em},
  hline{4} = {1-8}{},
  hline{6,8,10} = {1-8}{},
}
 \textbf{Setting}                    & \textbf{Five Most Popular Tags}&&&&& \textbf{TagSim(}$\mathbf{u_1,u_2}$\textbf{)} \\
{Source Domain\\Training History}                               & $u_1:$ Novelty & Jewelry & Sweatshirts & Running & Petite &  0.80             \\
   & $u_2:$ Novelty & Jewelry & Sweatshirts & Running & Athletic  &                   \\
   \hline
{Target Domain\\Training History}   & $u_1:$ Accessories & Boot Shop & Women & Running & Petite  & 0.00                   \\
                                                   & $u_2:$ Fishing Gloves & Outdoor Gear & Men & Fishing & Hunting &                     \\
{Target Domain\\Testing Ground Truth  }  & $u_1:$Accessories & Pants & Women & Running & Fan Shop  & 0.20\\
& $u_2:$ Exercise & Fitness & Outdoor Gear & Camping & Fan Shop  & \\
\hline
 {Target Domain Results \\ by UniCDR}        & $u_1:$ \textcolor{red}{Accessories} & Socks & \textcolor{red}{Women} & \textcolor{red}{Running} & Petite  & 0.60            \\
                                                & $u_2:$ Accessories & Snowshoes & Women & Running &  Fishing &                     \\
{Target Domain Results \\ by CUT-LightGCN}  & $u_1:$ \textcolor{red}{Accessories} & \textcolor{red}{Pants} & Team Sports & \textcolor{red}{Running} & Sunglasses  & 0.00                   \\
                                                 & $u_2:$ Gun Storage\&Safes	&	\textcolor{red}{Outdoor Gear} & \textcolor{red}{Camping} &	First Aid Kits&Nets&
\end{tblr}
\label{fig:case}
\end{table*}

\subsection{The Negative Transfer Problem (RQ2)}
Comparing the results of single domain methods trained on the target domain and both domains, we find that negative transfer indeed happens. For, training on both domains will lead to worse performance on Amazon Cloth and Video, as well as Douban Movie and Music as shown in Table \ref{tab:Experiment}.
For Amazon Cloth and Video, it results from the far domain relationship.
For Douban, especially Douban Movie, the source \& target domain datasets have less sparsity, where a single domain model is already enough for the target dataset, and a dense source-domain dataset may induce more source-domain exclusive collaborative knowledge. From Table \ref{tab:Experiment}, introducing source domain collaborative information without filtering degrades target performance in the above scenarios. Our CUT framework significantly outperforms the corresponding single-domain backbone trained on both domains, which indicates that CUT can precisely identify useful source knowledge based on the target user similarity.
Generally, existing cross-domain baselines outperform single-domain methods, especially the DTCDR and UniCDR. However, neglecting the user similarity relationship shift across domains still undermines their performance, especially when the source dataset provides less useful information. 

The first part of the \textbf{Answer to RQ2}: Experimental results empirically verify that our framework is effective towards the negative transfer problem, CUT performs well both when the source dataset is closely related to the target dataset or otherwise.

\subsection{Negative Transfer Case Study~(RQ2)}
We further conduct a case study demonstrating how our CUT framework addresses the negative transfer issue by maintaining target user similarity~(\textbf{RQ2}). 
As shown in Table \ref{fig:case}, we choose two overlapping users $u_1$ and $u_2$ from the source domain AmazonCloth and the target domain AmazonSports. 
We represent users' preferences with the five most popular item tags from all their historical interactions, 
and similarity between users is calculated by the overlapping ratio of the 5 tags. We also obtain the five most popular item tags from the top ten items generated by the recommendation model to represent model-predicted user preference.
The training history indicates that $u_1$ and $u_2$ share similar interests in the source domain but have distinct preferences in the target domain.
The strongest cross-domain baseline UniCDR transfers the user similarity in the source domain to the target domain, leading to inaccurate similar predictions in the target domain.
In contrast, our CUT Framework can keep the target-domain user similarities from being affected by the irrelevant source-domain user collaborative information. The item tags in the result list are colored red if it is also included in the testing ground truth. From Table \ref{fig:case}, UniCDR generates similar items for $u_1$ and $u_2$ based on their similar preference in the source domain. However, its lack of filtering irrelevant source knowledge partially sacrifices the accuracy of $u_2$ in the target domain.

Second part of the \textbf{Answer to RQ2}:
The case study shows that compared to UniCDR, the CUT framework is more robust to irrelevant source-domain user collaborative information.

\subsection{Ablation Studies on TRANSFER phase~(RQ3)}
\begin{figure}[hbt]
    \centering
    \includegraphics[width=0.9\linewidth]{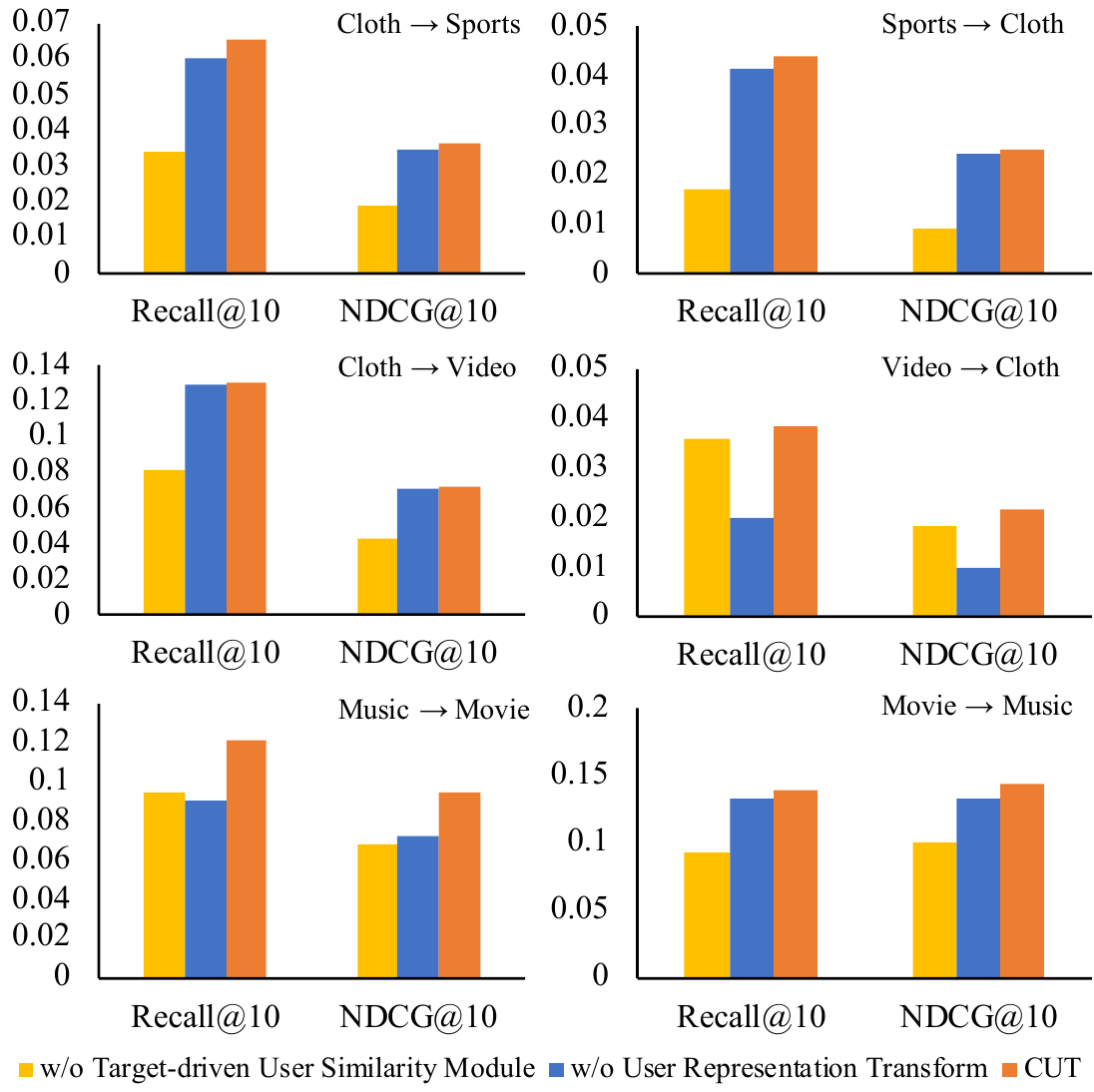}
    \caption{Ablation study of the target-driven user similarity module and user representation transformation module in TRANSFER phase.}
    \label{fig:ab}
\end{figure}

To answer \textbf{RQ3}, we conduct ablation studies 
on all six CDR tasks in Figure \ref{fig:ab}. Specifically, we compare the target domain performance of the following variants of our CUT framework:\\
\textbf{CUT w/o target driven user similarity module}: CUT without the guidance from the user similarity-based loss term $\mathbf{L}^c$.\\
\textbf{CUT w/o user representation transform}: CUT without the user representation transformation layer. Then the same user embedding is used to represent user behaviors in separate domains. \\
\textbf{CUT}: The complete version of the CUT framework that adopts the LightGCN as the single-domain backbone.

In most cases, the lack of contrastive loss term will cause a severe drop in performance, which is even worse than simply training an MF model on the target domain. This indicates that the framework can not learn suitable addition parameters in the user transformation layer for the target domain without additional supervision signals to amplify the target-domain user similarity relationships.
The CUT framework without the transform layer shares its user representation across domains. In other words, overlapped users do not have double versions of user representations as described in Section \ref{sec:rep_trans}. The lack of the transformation layer $\mathbf{F}$ results in a limited capacity for modeling user behaviors. This forces the model to learn a user embedding that is suitable for both domains, which is sub-optimal because users behave differently across domains naturally. 
However, the contrastive loss term assures that the model remembers the target-domain user similarity relationships. Therefore, its performance drops slightly in most cases.

\subsection{Backbone Model-based User Similarity Learning in TARGET phase~(RQ3)}
\label{sec:exp_similarity}
\begin{figure}[hbt]
    \centering
    \includegraphics[width=0.9\linewidth]{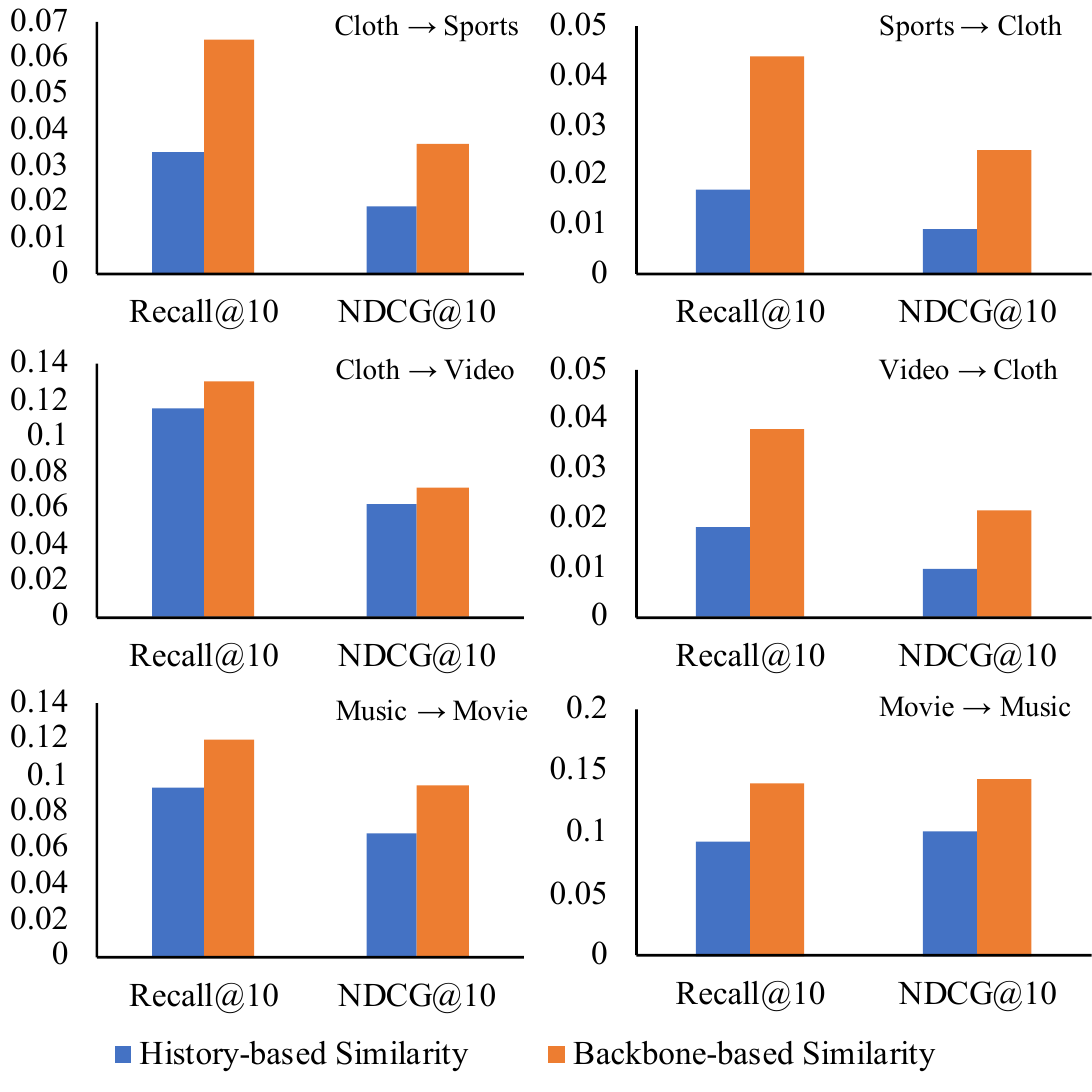}
    \caption{Performance comparisons of the user history-based similarity and the backbone-driven similarity.}
    \label{fig:sim}
\end{figure}

For the TARGET phase,
we illustrate how the user similarity matrix contributes to CDR tasks by comparing the backbone-driven~(Eq.\ref{eq:train_sim}) and history-driven~(Eq.\ref{eq:his_sim}) similarity~(\textbf{RQ3}). 
As shown in Figure \ref{fig:sim}, the similarity matrix derived by the user embeddings trained by the backbone model on the target dataset consistently achieves better results than the history-based similarity matrix on six CDR tasks. 
As we explained in Section~\ref{seq:sim}, simply calculating cosine distance based on one-hot user interaction history yields a static similarity matrix for different single-domain backbones, which leads to sub-optimal user similarity relationship modeling than backbone-adaptive ones.

From the above two subsections, we have \textbf{Answer to RQ3}: The guidance of target-driven user similarity in CUT is crucial during the learning process of cross-domain user representation. Furthermore, the user similarity that corresponds with the backbone recommendation model achieves better performance than the simple history-based user similarity.
\section{Conclusion}
In this paper, we provide a novel perspective on the negative transfer issue in cross-domain recommendation~(CDR) tasks, i.e., the distortion of user similarity relationships. 
Then we propose a Collaborative information regularized User Transformation~(CUT) framework to alleviate negative transfer by directly filtering source-domain user collaborative information with target-domain user similarity constraints.
In CUT, a two-phase training process is adopted to learn user similarities in the target domain~(TARGET phase), and then selectively transfer useful information from the source domain~(TRANSFER phase).
Specifically, we design a user transformation layer and a contrastive loss to constrain the representations of overlap users, which help maintain the user relationships in the target domain when introducing source-domain information.
We conduct extensive experiments on real-world datasets, where CUT shows significant improvements to backbone models, as well as compared with state-of-the-art single and cross-domain baselines.
Further comparisons and analyses also illustrate that CUT effectively filters out irrelevant source user collaborative information, and thus successfully alleviates the negative transfer issue.
As a framework, CUT enhances the cross-domain performance for various single-domain recommenders, while keeping their model structures and loss terms.
We believe there will be more attempts to apply single-domain recommendation models for CDR tasks.

\clearpage
\bibliographystyle{ACM-Reference-Format}
\bibliography{sample-authordraft}
\nocite{*}
\appendix

\end{document}